\documentclass[aps,pra,twocolumn,groupedaddress,showpacs]{revtex4-2}

\usepackage{graphicx,dcolumn,bm,xstring,braket,amsmath}
\usepackage[unicode=true,pdfusetitle,bookmarks=true,bookmarksnumbered=true,bookmarksopen=false, breaklinks=false,pdfborder={0 0 0},pdfborderstyle={},backref=false,colorlinks=false, pdftitle={Experimental Determination of the D1 Magic Wavelength for 40K}]{hyperref}

\usepackage{float}

\usepackage{xcolor}  

\usepackage{tikz}
\usepackage{pgfplots}
	\pgfplotsset{compat=1.18} 
\usetikzlibrary{calc, decorations.pathmorphing, shapes.geometric, arrows.meta}

\hypersetup{
    colorlinks=true,
    citecolor=blue    
}

\begin{document}

\title{Experimental Determination of the $D1$ Magic Wavelength for $^{40}$K}

\author{Guy Hay Kalifa, Dor Kopelevitch, Amir Stern and Yoav Sagi}

\email[Electronic address: ]{yoavsagi@technion.ac.il}

\affiliation{Physics Department and Solid State Institute, Technion -- Israel Institute of Technology, Haifa 32000, Israel}
\date{\today} 

\begin{abstract}
Neutral-atom arrays offer a promising path for quantum simulation, yet the potential of fermionic $^{40}$K remains largely constrained by state-dependent light shifts that degrade cooling and detection fidelities. This problem can be resolved by working at a magic wavelength, where the differential light shift vanishes. We report the first experimental determination of the magic wavelength for the D1 transition in fermionic $^{40}$K at 1227.54(4) nm. Using in-trap loss spectroscopy in a wavelength-tunable optical tweezer, we map the differential AC Stark shift across a range of trapping powers and wavelengths. By converting these shifts to differential scalar polarizabilities, we find excellent agreement with relativistic all-order calculations. Benchmark measurements at 1064.49 nm further quantify the intensity-sampling systematics inherent to standard trapping wavelengths, contrasting with the ``mechanically clean'' environment provided by the magic condition. Our results provide an important step toward high-fidelity in-trap D1 cooling, fluorescence imaging, and light-assisted loading, establishing a robust path toward scaling fermionic neutral-atom arrays for quantum information science.
\end{abstract}

\maketitle

\section{Introduction}
Ultracold atomic gases have been instrumental in exploring quantum many-body physics \cite{many_body_1,many_body_2,many_body_3}, quantum simulation and computation \cite{quantum_simulation_1,quantum_simulation_2,quantum_simulation_3,quantum_simulation_4}, and high-precision metrology \cite{metrology1,metrology2}. The interaction of atoms with far-detuned light allows one to exert repulsive or attractive forces and trap atoms in flexible geometries \cite{Grimm2000}. Among these, optical tweezers have emerged as a particularly powerful tool, enabling the deterministic preparation, manipulation, and detection of small ensembles, down to a single atom, with sub-micron spatial resolution~\cite{Manetsch2025,tweezers2,tweezers3,tweezers4,tweezers5,tweezers6}. Consequently, optical tweezers have assumed a central role in neutral-atom-based quantum computation \cite{Graham2022,Evered2023,Bluvstein2024,Bluvstein2025,Manetsch2025,Chung2025}, the implementation of tight-binding models \cite{Kaufman2014,Murmann2015,Spar2022,Young2022,Florshaim2024,Eckner2025,Jain2025}, and advanced quantum sensing \cite{Norcia2019,Madjarov2019,Young2020,Nemirovsky2023,Meltzer2024,Schaeffner2024,Finkelstein2024}.

While tweezer technology is highly advanced for bosonic species, scaling fermionic arrays, such as those using $^{40}$K, remains a major frontier. A significant bottleneck is the AC Stark shift induced by the trapping light. For alkali atoms in linearly polarized optical fields, and specifically for states with total electronic angular momentum $j=1/2$, the vector and tensor contributions to the dynamic polarizability vanish, and the light shift is fully described by the scalar polarizability. In this case, the AC Stark shift of an atomic level $\mu$ is given by \cite{magic_alkaly,Magic_wavelengths_2013,vector_polarizability_2}:
\begin{equation}
	\Delta E_\mu = -\frac{\alpha_0^{\mu}(\omega)}{2\varepsilon_0 c}\,I,
	\label{eq:stark}
\end{equation}
where $\alpha^{\mu}_0(\omega)$ is the scalar dynamic polarizability at the trap laser angular frequency $\omega$, $I$ is the local intensity of the trapping light, $\varepsilon_0$ is the vacuum permittivity, and $c$ is the speed of light \cite{Grimm2000}. The scalar polarizability is expressed as a sum over all dipole-coupled excited states $\nu$:
\begin{equation}
	\alpha_0^\mu(\omega) = \frac{2}{3(2j_\mu+1)}
	\sum_{\nu}\frac{|\langle \nu||d||\mu\rangle|^{2}\,(E_\nu-E_\mu)}
	{(E_\nu-E_\mu)^{2}-(\hbar\omega)^{2}},
\end{equation}
where $(E_\nu-E_\mu )$ is the transition energy between $|\mu\rangle$ and $|\nu\rangle$, and $\langle \nu||d||\mu\rangle$ is the reduced dipole matrix element.

\begin{figure*}[t]
    \centering
    \resizebox{\textwidth}{!}{
    \begin{tikzpicture}[
        font=\sffamily\small,
        >=Stealth,
        level/.style={thick, black},
        virtual/.style={thick, dashed, gray},
        photon/.style={->, decorate, decoration={snake, amplitude=1pt, segment length=5pt}, thick},
        trap/.style={thick, color=#1, smooth, samples=50, domain=-1.5:1.5}
    ]

    \node[anchor=north west, font=\bfseries\normalsize] at (0, 5.5) {(a)};
    
    \begin{scope}[shift={(0.8, 1)}]
        \draw[level] (0, 0) -- (1.5, 0) node[right, black] {$4^2S_{1/2}$};
        
        \draw[virtual] (0, -0.8) -- (1.5, -0.8);
        \draw[<->, shorten >=1pt, shorten <=1pt, blue] (1.2, 0) -- (1.2, -0.8) node[midway, right] {$\Delta E_g$};

        \draw[level] (0, 3) -- (1.5, 3) node[right, black] {$4^2P_{1/2}$};
        
        \draw[virtual] (0, 2.0) -- (1.5, 2.0);
        \draw[<->, shorten >=1pt, shorten <=1pt, red] (1.2, 3) -- (1.2, 2.0) node[midway, right] {$\Delta E_e$};

        \draw[photon, orange!80!red] (-1.2, -0.4) -- (0, -0.4) node[midway, above, scale=0.8] {Tweezer};
        \draw[photon, orange!80!red] (-1.2, 2.5) -- (0, 2.5) node[midway, above, scale=0.8] {Tweezer};

        \draw[->, thick, purple] (0.5, -0.8) -- (0.5, 2.0) node[midway, right] {$770$ nm Probe};
    \end{scope}

    \node[anchor=north west, font=\bfseries\normalsize] at (4, 5.5) {(b)};
    
    \begin{scope}[shift={(6.2, 0.5)}]
        \node[above] at (0, 4.3) {$\lambda \neq \lambda_m$};
        \draw[->] (-1.8, -1.0) -- (-1.8, 4.2) node[left] {$E$};
        \draw[->] (-1.8, -1.0) -- (1.8, -1.0) node[below] {$x$};
        
        \draw[trap=blue] plot (\x, {-0.5*exp(-2*\x*\x)}); 
        \draw[trap=red] plot (\x, {1.5 + 2.63*exp(-2*\x*\x)}); 
        
        \node[blue, right] at (1.2, 0.15) {$U_g$};
        \node[red, right] at (1.2, 1.8) {$U_e$};
    \end{scope}

    \begin{scope}[shift={(10.2, 0.5)}]
        \node[above] at (0, 4.3) {$\lambda = \lambda_m$};
        \draw[->] (-1.8, -1.0) -- (-1.8, 4.2) node[left] {$E$};
        \draw[->] (-1.8, -1.0) -- (1.8, -1.0) node[below] {$x$};
        
        \draw[trap=blue] plot (\x, {-0.395*exp(-2*\x*\x)}); 
        \draw[trap=red] plot (\x, {1.5 - 0.395*exp(-2*\x*\x)}); 
        
        \node[blue, right] at (1.2, 0.15) {$U_g$};
        \node[red, right] at (1.2, 1.8) {$U_e$};
    \end{scope}

    \node[anchor=north west, font=\bfseries\normalsize] at (12.5, 5.5) {(c)};
    
    \begin{scope}[shift={(14.2, 2)}]
        \filldraw[thick, fill=cyan!10, draw=cyan!80!black] (-1, 0) ellipse (0.15 and 1.8);
        \node[above, scale=0.9] at (-1, 1.8) {Objective};
        
        \fill[orange!50!red, opacity=0.3] 
            (-0.91, 1.5) arc (90:-90:0.08 and 1.5) -- (2, -0.1) -- (2, 0.1) -- cycle;
        
        \fill[orange!50!red, opacity=0.3] (2, 0.1) -- (2.8, 0.3) -- (2.8, -0.3) -- (2, -0.1) -- cycle;
        
        \node[orange!80!red, align=center, scale=0.9, font=\bfseries] at (0.3, 0) {Tweezer\\$1227$ nm};
        
        \shade[ball color=blue] (2, 0) circle (0.15);
        \shade[ball color=blue] (1.85, 0.08) circle (0.12);
        \shade[ball color=blue] (2.1, -0.05) circle (0.1);
        \node[right, scale=1] at (2.1, 0) {$^{40}$K};
        
        \draw[photon, purple] (2, -1.8) -- (2, -0.4);
        \node[below, purple, scale=0.8, align=center] at (2, -1.8) {$D1$ Probe\\\& Repump};

        \draw[->, purple, thick] (2.15, -1.4) arc (0:330:0.15 and 0.04);
        \node[right, purple, scale=1, font=\bfseries] at (2.1, -1.4) {$\sigma^+$};

        \draw[->, thick, black] (1.1, -1.6) -- (1.1, -0.8) node[midway, left] {$B$};
        
        \draw[photon, green!60!black] (1.8, 0.4) -- (1.4, 1.2);
        \draw[photon, green!60!black] (2.2, 0.4) -- (2.6, 1.2);
        \draw[photon, green!60!black] (1.8, -0.4) -- (1.4, -1.2);
        \draw[photon, green!60!black] (2.2, -0.4) -- (2.6, -1.2);
        \node[above, green!60!black, scale=0.9] at (2, 1.2) {Fluorescence};
    \end{scope}

    \end{tikzpicture}
    } 
    
    \caption{\textbf{Principles of the experiment.} (a) Level scheme of the $D1$ transition ($4^2S_{1/2} \rightarrow 4^2P_{1/2}$) in $^{40}$K. The tweezer field induces state-dependent AC Stark shifts ($\Delta E_g, \Delta E_e$). (b) Exact spatial profile of the trapping potentials $U(x) = -\alpha_0 I(x)/(2\varepsilon_0 c)$ for an off-magic $\lambda \neq \lambda_m$ tweezer (left) versus the magic wavelength $\lambda = \lambda_m$ (right). For the off-magic wavelength, we used the polarizabilities at $1064$~nm, where the strongly repulsive excited state creates a large differential shift compared to the attractive ground state. At the magic condition, the identical potentials ensure a mechanically clean, spatially uniform resonance. (c) Simplified experimental geometry showing the high-NA objective, the near-resonant $D1$ probe beam propagating perpendicular to the tweezer axis, and the fluorescence emission, which eventually leads to atom loss. We measure the number of atoms by releasing them from the tweezer, recapturing them in a 3D MOT, and recording their fluorescence signal for 500~ms.}
    \label{fig:principle}
\end{figure*}
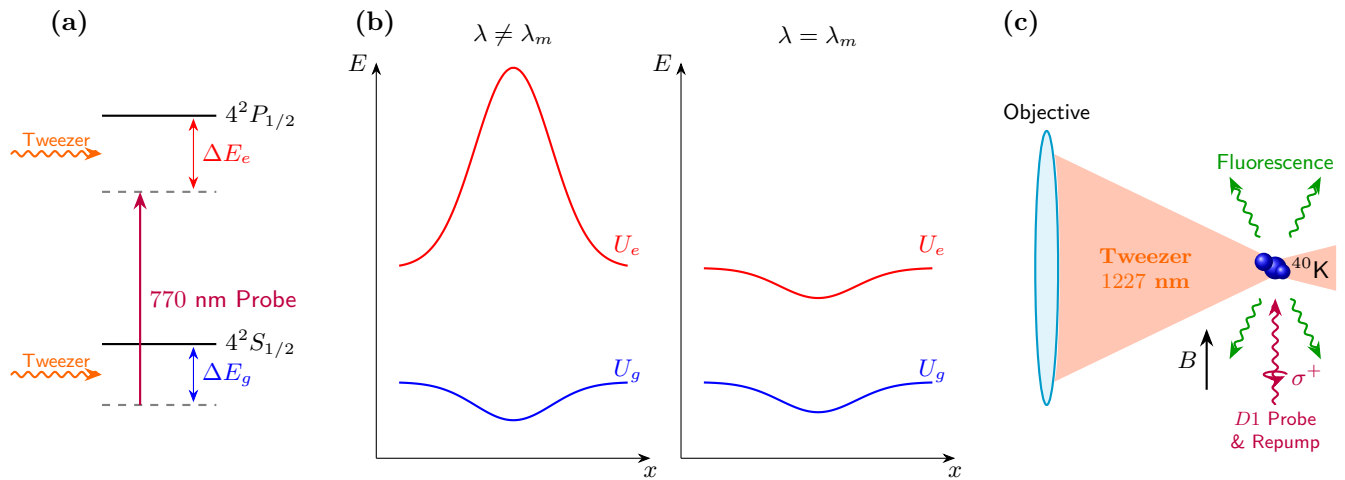

Differences in polarizability between states result in a differential AC Stark shift, leading to power-dependent frequency shifts and spectral broadening that degrade the fidelity of spectroscopy, cooling, and detection  (Figure~\ref{fig:principle}b). These effects are eliminated at a \textit{magic wavelength}, where the polarizabilities of the ground and excited states are equal \cite{magic_alkaly}. While magic wavelengths have been measured for several alkali species, including Rb \cite{Lundblad_2010}, Cs \cite{Liu_2017}, and Na \cite{d1_magic_sodium}, the magic wavelength for the $D1$ transition of $^{40}$K has, until now, only been predicted theoretically \cite{Magic_wavelengths_2013,2025Portal}; the most recent calculation gives $\lambda_m^{\text{theory}}=1227.55~\text{nm}$ \cite{2025Portal}. An experimental benchmark is therefore essential for high-fidelity control in potassium-based arrays.

In this Letter, we report the first experimental determination of the $D1$ magic wavelength in $^{40}$K. Using loss spectroscopy on few-atom ensembles in a wavelength-tunable optical tweezer, we measure the transition frequency shift as a function of trap power and identify the zero crossing of the differential Stark shift. Our measured value, $\lambda_m=1227.54(4)~\text{nm}$, confirms theoretical predictions with high precision. Operating at the magic wavelength can provide several technical advantages crucial for large-scale $^{40}$K arrays, including high-fidelity fluorescence detection under unshifted resonance conditions, improved efficiency of $D1$-based light-assisted loading, and the ability to apply $D1$ gray-molasses cooling directly inside the tweezer.

\section{Measurement Procedure}
The determination of the magic wavelength relies on measuring the differential AC Stark shift through in-trap loss spectroscopy (Figure~\ref{fig:principle}c). Our approach employs a wavelength-tunable optical tweezer loaded with a small ensemble of $^{40}$K atoms prepared in a well-defined hyperfine ground state. We subject the trapped atoms to a short, near-resonant $D1$ probe pulse; when the pulse is close to resonance with the light-shifted transition, the resulting photon scattering and subsequent heating lead to the ejection of atoms from the trap. By scanning the probe frequency and recording the remaining atom number, we extract the transition frequency shift as a function of the tweezer power. This procedure is repeated across a range of trapping wavelengths to determine the differential Stark shift per unit power, allowing us to identify the magic wavelength as the zero crossing where the ground and excited state polarizabilities are precisely matched.

\textit{Experimental sequence.}---The experimental cycle begins with the loading of a magneto-optical trap (MOT), followed by $D1$ gray molasses cooling to prepare $N\approx5\times 10^6$ atoms at $T\approx5~\mu\text{K}$. Degenerate Raman sideband cooling \cite{Elad_thesis} further reduces the temperature to $T\approx1~\mu\text{K}$. Microwave spectroscopy verifies that the resulting ensemble is spin-polarized, with $80\%$ of the population in the $\ket{9/2, -9/2}$ state and $20\%$ in $\ket{9/2, -7/2}$. The atoms are then transferred to a $1064~\text{nm}$ crossed optical dipole trap, where evaporative cooling produces a sample of $N\approx3\times 10^4$ atoms at $T\approx10~\mu\text{K}$.

Subsequently, approximately $100$ atoms are loaded into a linearly polarized optical tweezer with a tunable wavelength between $1226$ and $1229~\text{nm}$. In-trap optical evaporation is performed by ramping the tweezer power to $P=5.8~\text{mW}$, leaving $N\approx50$ atoms at $T\approx12(1)~\mu\text{K}$, measured by the release-and-recapture method \cite{Tuchendler2008}. The power is then ramped adiabatically to the target values used for spectroscopy.

The tweezer light is generated by a DFB laser diode (Innolume DFB-122700-PM-050-MFV) seeded into a semiconductor optical amplifier (Innolume SOA-1250-110-PM-27-DB), providing over 100~mW of power. To ensure absolute wavelength accuracy, we calibrated our spectrum analyzer using a reference laser locked to the $^{40}$K $D1$ transition via saturated absorption spectroscopy (SAS). This calibration was further cross-validated with a high-precision wavemeter, showing agreement within $0.01$ nm. Stability measurements confirm the trapping wavelength remains within $0.02$ nm over 14 hours of operation. 

The beam is focused through a high-NA objective ($\mathrm{NA}=0.75$) designed for diffraction-limited performance at $1064$~nm. Characterization via trap-frequency parametric-excitation measurements at $1227$~nm reveals a Gaussian waist of $w_0 = 1.6(1)~\mu\text{m}$. This departure from the diffraction limit is attributed to a predicted r.m.s.\ wavefront error of $\approx 0.14\lambda$ at the longer wavelength. 

\begin{figure}[t]
    \centering
   
    \includegraphics[width=1\linewidth]{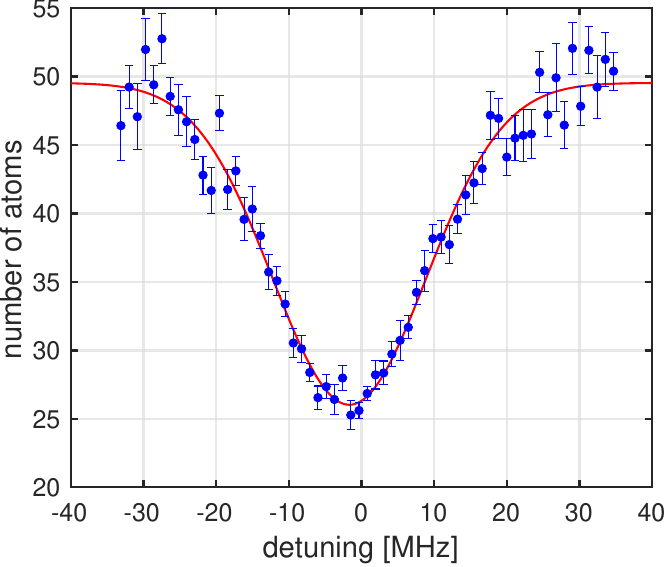}
    
    \caption{\textbf{Atom-loss spectroscopy of the $D1$ transition in an optical tweezer.} Measured atom number remaining in the trap following a $10~\mu\text{s}$ probe pulse as a function of probe detuning. For this representative trace, the trapping wavelength is $1227.78$~nm and the power is $5.8$~mW. The detuning is defined relative to the unshifted $D1$ resonance frequency in free space, which is independently calibrated by performing spectroscopy on an untrapped atomic cloud with the tweezer light and all magnetic trapping fields extinguished. The solid red line is a Gaussian fit to the loss spectrum. The extracted center is $\Delta \nu=-1.7(2) ~\text{MHz}$, which determines the light-shifted transition frequency, and the width is $W=10.6(3)~\text{MHz}$. Error bars represent the standard error of the mean from 8 repeated experimental cycles.} 
    \label{fig:1227.78nm_10mW}
\end{figure}

Loss spectroscopy of the $D1$ transition is performed using a circularly polarized probe beam with an intensity equal to the saturation intensity, $I_{\text{sat}}=1.70~\text{mW/cm}^2$. The probe detuning $\Delta$ is defined relative to the $4^2S_{1/2}\ket{F=9/2} \rightarrow 4^2P_{1/2}\ket{F'=7/2}$ transition. A repump beam addressing the $4^2S_{1/2}\ket{F=7/2} \rightarrow 4^2P_{1/2}\ket{F'=9/2}$ transition is applied simultaneously to prevent population accumulation in dark states. A uniform magnetic field of $B=4$~G, aligned along the probe's propagation direction, lifts the $m_F$ degeneracy. When resonant with the light-shifted transition, the probe induces photon scattering and heating that ejects atoms from the trap within $10~\mu\text{s}$. The remaining atom number is recorded via fluorescence imaging, by recapturing the atoms in the MOT \cite{Shkedrov2018}. A typical result of such a measurement is shown in Figure~\ref{fig:1227.78nm_10mW}. 

We fit the resonance signal with a Gaussian and extract both the resonance detuning and linewidth. The extracted linewidths lie in the range of $10$–$14$~MHz, consistent with expectations. Power broadening at saturation intensity increases the natural linewidth to approximately $8.5$~MHz, with an additional $\sim2.5$~MHz contribution coming from Zeeman splitting in a $4$~G magnetic field. The focusing-induced vector light shift discussed below contributes further broadening: the shift varies linearly across the thermal distribution and, being odd in position, broadens the line symmetrically without displacing its center, with an r.m.s.\ magnitude growing from $\sim3$~MHz at the lowest tweezer powers to $\sim5$~MHz at the highest. Doppler broadening, which increases the observed width by only $\sim1$~kHz,  can be safely neglected. 

\section{Results and Discussion}
\begin{figure}[t]
	\centering
	\includegraphics[width=1\linewidth]{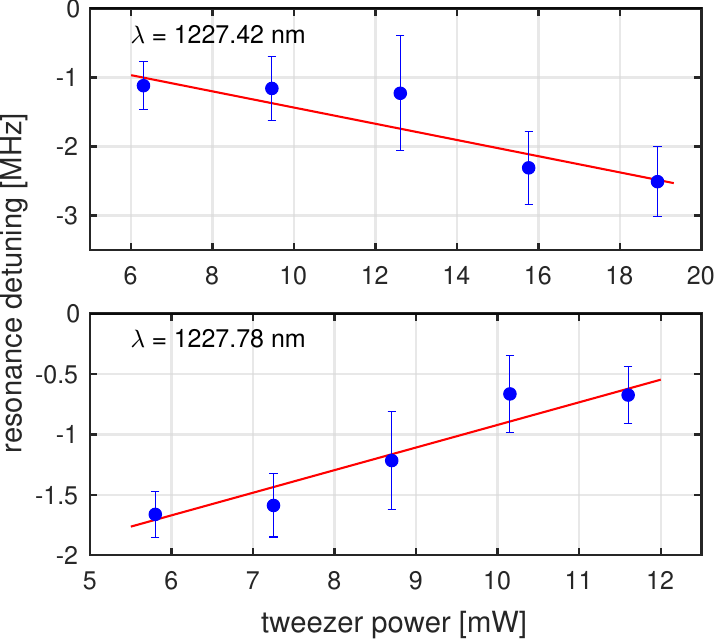}
	\caption{\textbf{Linear scaling of the differential AC Stark shift.}
		Differential light shift $\Delta \nu$ of the $D1$ resonance as a function of
		optical tweezer power for trapping wavelengths on either side of the magic
		wavelength: $\lambda=1227.42$~nm (upper panel), where the slope is negative,
		and $\lambda=1227.78$~nm (lower panel), where it is positive. The shift is
		defined relative to the unshifted free-space resonance ($\Delta \nu = 0$).
		Weighted linear fits (solid lines) yield $\Delta \nu /P = -0.12(2)$ and
		$+0.19(3)$~MHz/mW, respectively. Error bars represent the standard error of the mean.}
	\label{fig_1227_slope}
\end{figure}

For each tweezer wavelength and final trap power, we performed loss spectroscopy measurements and extracted the light-shifted resonance transition detuning, $\Delta \nu$, relative to the free-space $D1$ resonance. Examples of such scans for tweezer wavelengths on either side of the magic wavelength are shown in Figure~\ref{fig_1227_slope}. By repeating this measurement at several final tweezer powers, we determined the differential light shift per unit power, $\Delta \nu / P$. For all wavelengths studied, the shift $\Delta\nu$ exhibited a clear linear dependence on the tweezer power, confirming that the measurement is performed in a regime dominated by the scalar dynamic polarizability.

The fitted lines have intercepts of $-0.3(5)$ and $-2.8(4)$~MHz for the upper
and lower panels of Fig.~\ref{fig_1227_slope}, respectively. Such offsets are
expected due to the way we perform the calibration and measurement. The free-space reference is recorded at zero magnetic field on an unpolarized cloud, whereas the in-trap spectroscopy is performed at $B=4$~G on the open $\sigma^{+}$ transition $F=9/2\rightarrow F'=7/2$, whose Zeeman components extend from $+7.1$~MHz (for $m_F=-9/2$) to $-4.6$~MHz ($m_F=+5/2$). The measured offsets lie within this range, and their precise values are sensitive to how the initial population is distributed over the $m_F$ sublevels and how it evolves under optical pumping during the pulse. The same Zeeman structure contributes $\sim2.5$~MHz to the observed linewidths. Importantly, these offsets are independent of the tweezer power and affect neither the extracted slopes nor the determination of $\lambda_m$.

\begin{figure}[t]
    \centering
    \includegraphics[width=1\linewidth]{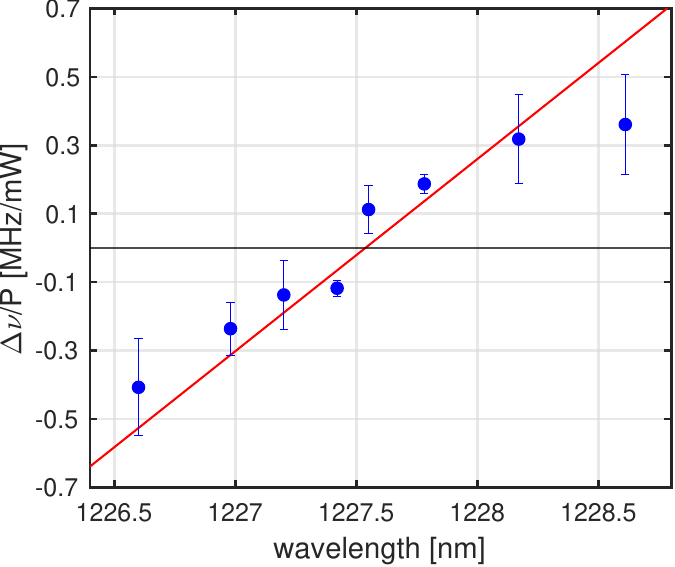}
    \caption{\textbf{Determination of the $D1$ magic wavelength for $^{40}$K.} The differential AC Stark shift per unit power, $\Delta \nu / P$, is plotted as a function of the trapping wavelength. The magic wavelength is identified as the zero crossing of a linear fit (solid line), representing the point where the scalar polarizabilities of the ground and excited states are precisely matched. Error bars denote the statistical uncertainty from the individual slope extractions.}
    \label{fig_magic_plot}
\end{figure}

The extracted slopes, $\Delta \nu / P$ (in units of MHz/mW), are plotted as a function of the tweezer wavelength in Figure~\ref{fig_magic_plot}. The data exhibit a linear dependence within the scanned range, consistent with the expectation that the differential Stark shift changes sign near the magic wavelength. A linear fit with two free parameters is applied to the data, and the zero crossing of the fit is identified as the magic wavelength. From this analysis, we determine the magic wavelength to be $\lambda_m=1227.54(4)~\text{nm}$, in excellent agreement with the theoretical prediction of $\lambda_m^{\text{theory}}=1227.55~\text{nm}$ \cite{2025Portal}. The contributions to the total experimental uncertainty are summarized in Table~\ref{tab:uncertainty}.

\begin{table}[b]
	\caption{\label{tab:uncertainty}Uncertainty budget for the $D1$ magic wavelength
		determination. Multiplicative systematic effects (optical power calibration, trap
		geometry) rescale all slopes uniformly and therefore do not displace the zero
		crossing; they are discussed in the text. The total uncertainty is the
		root-sum-square of the individual contributions.}
	\begin{ruledtabular}
		\begin{tabular}{lc}
			Source of uncertainty & Value (nm) \\
			\hline
			Laser wavelength calibration  & 0.02 \\
			Wavelength stability (14-hour drift) & 0.02 \\
			Statistical uncertainty (linear-fit zero crossing) & 0.03 \\
			Residual vector light shifts & $<0.01$ \\
			Hyperfine dependence of $\lambda_m$ & $<0.001$ \\
			\hline
			\textbf{Total uncertainty} & \textbf{0.04} \\
		\end{tabular}
	\end{ruledtabular}
\end{table}

\begin{figure}[t]
    \centering
    \includegraphics[width=1\linewidth]{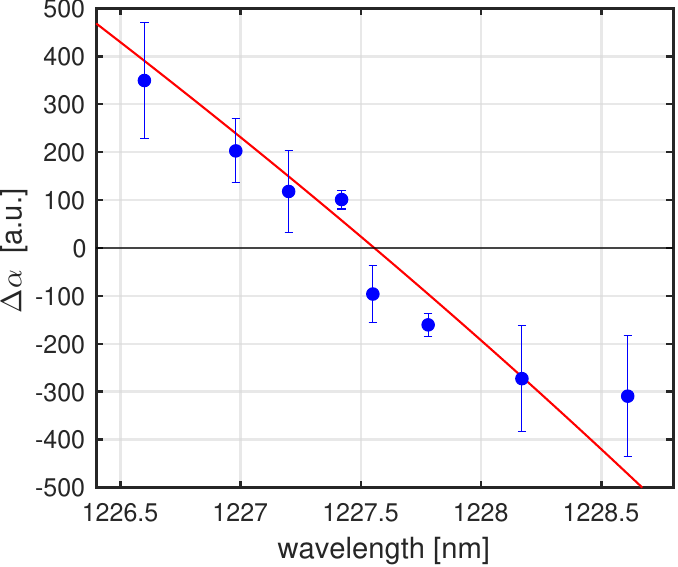}
   \caption{\textbf{Experimental determination of the $D1$ differential
   		polarizability.} The measured differential scalar polarizability
   	$\Delta\alpha$ (points) is plotted as a function of the trapping wavelength.
   	The data are extracted from the light-shift slopes using the measured beam
   	waist and compared to a theoretical curve (solid line) obtained from
   	relativistic all-order calculations \cite{2025Portal}. The conversion carries
   	an overall $\pm13\%$ scale uncertainty from the $w_0^{2}$ dependence.}
    \label{fig_magic_plot_mK}
\end{figure}

\textit{Systematic effects.}---The magic wavelength is identified as the zero
crossing of $\Delta\nu/P$ versus $\lambda$, which makes the measurement inherently
robust against most systematic effects. Multiplicative errors that rescale all
slopes by a common factor, including uncertainties in the optical power calibration, in the
beam waist and trap geometry, or in the wavelength dependence of the photodiode
response (smooth at the sub-percent level over the $3$~nm scan), leave the zero
crossing unchanged; they affect only the vertical scale of
Fig.~\ref{fig_magic_plot} and enter the polarizability comparison of
Fig.~\ref{fig_magic_plot_mK} through the $w_0^2$ conversion factor ($\pm13\%$ scale
uncertainty). Constant, power-independent frequency offsets, such as the Zeeman shift of
the probed transitions at $B=4$~G, AC Stark shifts induced by the probe and repump
light themselves ($\lesssim 30$~kHz,  no correction is required), and photon-recoil
Doppler shifts, are absorbed into the intercepts of the $\Delta\nu(P)$ fits and do
not enter the slopes. Magnetic-field calibration and drift therefore do not
propagate to $\lambda_m$. Only an additive contribution to the slope could displace the crossing. Given the fitted dispersion of $0.56$~MHz/(mW\,nm), a displacement of $\lambda_m$ by $0.01$~nm would require an
offset of $5.6$~kHz/mW. The largest such candidate, a residual vector light shift of
the tightly focused tweezer, is bounded below this level as discussed next.

For states with \(j=1/2\), the electronic tensor polarizability vanishes, whereas the vector polarizability does not. A sum-over-states evaluation using recommended potassium matrix elements
\cite{Magic_wavelengths_2013} gives $\alpha_v(4P_{1/2})\simeq1.60\times10^{4}$~atomic units (a.u., where $1~\mathrm{a.u.}=4\pi\varepsilon_0 a_0^{3}$ and $a_0$ is the Bohr radius) at \(1227.5\)~nm, approximately thirty times the scalar value. The enhancement results from the near cancellation of the scalar contributions \(+5.9\times10^{3}\) a.u. from \(4P_{1/2}\rightarrow3D_{3/2}\) and \(-5.2\times10^{3}\) a.u. from \(4P_{1/2}\rightarrow5S_{1/2}\); their vector contributions have the same sign and therefore add. The ground-state vector polarizability is only \(-2.8\) a.u. For the dominant \(\ket{9/2,-9/2}\rightarrow\ket{7/2,-7/2}\) transition, \(\langle J_z\rangle_{e}=7/18\), giving a differential vector coefficient of \(6.2\times10^{3}\) a.u. per unit circularity projected along the quantization axis.

Tight focusing renders the polarization elliptical away from the axis, with a
degree of circularity $C\simeq4x/(kw_0^2)$ along the direction of the transverse
polarization \cite{vector_polarizability_2}, i.e. $C=x/3.3~\mu$m in our trap. The
associated shift is not small in absolute terms, reaching $0.5$~MHz/mW at the
r.m.s.\ thermal radius $\sigma_x\approx0.2~\mu$m, but $C$ is \emph{odd} in $x$
and vanishes on axis. Therefore, it cancels upon averaging over the thermal distribution and
contributes to the observed linewidth rather than displacing the line center.
Because the tweezer polarization is perpendicular to the probe, the photon-recoil
displacement during the pulse is orthogonal to the direction in which $C$ varies
and does not break this cancellation. Displacing $\lambda_m$ by $0.01$~nm would require the mean sampled position to lie more than $2$~nm off axis along the polarization direction, i.e.\ an
asymmetry of $1\%$ of the thermal width. Since $C$ is odd about the optical
axis, such an offset can arise only from a static transverse force. The largest,
gravity, displaces the cloud by $g/\omega_r^{2}\approx0.2$~nm even in our
shallowest trap. We therefore include a bound of $<0.01$~nm in
Table~\ref{tab:uncertainty}. The power-dependent broadening implied by the same
mechanism ($\pm3$~MHz at $5.8$~mW) is consistent with the observed linewidths.

Hyperfine corrections to the polarizabilities scale as the ratio of the hyperfine splittings ($\sim1.3$~GHz for $4S_{1/2}$, $\sim0.16$~GHz for $4P_{1/2}$) to the detunings of the transitions dominating
$\alpha^{4P_{1/2}}$ near $1227.5$~nm ($4P_{1/2}\rightarrow5S_{1/2}$ and
$4P_{1/2}\rightarrow3D_{3/2}$, detuned by $\sim3$~THz and $\sim12$~THz). Hence, they enter at the $10^{-4}$ level. The corresponding variation of $\lambda_m$ between hyperfine components is below $10^{-3}$~nm, negligible at our precision.

\textit{Comparison with theory.}---To compare our results directly with theoretical models, we transform the measured slopes $\Delta\nu/P$ into the differential scalar polarizability
$\Delta\alpha\equiv\alpha_0^{4P_{1/2}}(\omega)-\alpha_0^{4S_{1/2}}(\omega)$, expressed in atomic units (a.u.). From Eq.~(\ref{eq:stark}), the shift of the transition frequency at the trap center is $h\,\Delta\nu=-\Delta\alpha\,I_0/(2\varepsilon_0 c)$ \cite{Grimm2000,vector_polarizability_2}, where $I_0=2P/(\pi w_0^2)$ is the peak intensity of a Gaussian beam of power $P$ and waist $w_0$. Converting to atomic units gives
\begin{equation}
	\Delta\alpha = -\frac{h c\, w_0^{2}}{4 a_0^{3}}\,\frac{\Delta\nu}{P}.
\end{equation}
As shown in Figure~\ref{fig_magic_plot_mK}, the experimental values show excellent agreement with relativistic all-order calculations \cite{2025Portal}. The consistency between the measured polarizabilities and the theoretical curve, achieved without any fitting parameters, provides a robust cross-validation of our trap calibration and the underlying atomic matrix elements.

Benchmark measurements were performed at $1064.49$~nm to validate the
spectroscopic procedure at a standard trapping wavelength, where the differential
shift is large. At this wavelength the tweezer is slightly asymmetric;
parametric-resonance measurements of the radial trap frequencies yield waists of
$w_x=2.43(6)~\mu$m and $w_y=1.88(4)~\mu$m. 
From the loss spectroscopy we
extracted a differential light-shift slope of $2.06(7)$~MHz/mW, an order of
magnitude larger than at $1227.78$~nm (Figure~\ref{fig_1227_slope}). As
illustrated in Figure~\ref{fig:principle}b, this large shift arises from the
strongly repulsive $4^2P_{1/2}$ potential at $1064$~nm
($\alpha_P\approx-3155$~a.u.), which contrasts with the attractive ground-state
potential ($\alpha_S\approx599$~a.u.) \cite{2025Portal}. The corresponding
trap-center prediction is $2.45(9)$~MHz/mW. The measured slope is slightly
smaller because the thermal ensemble samples intensities below the peak. At the
temperatures of this data set, the thermal average of the Gaussian profile is
$\langle I\rangle/I_0\approx0.8$, giving a predicted effective slope of
$2.0(1)$~MHz/mW, in agreement with the measurement. We have also performed a semiclassical Monte Carlo simulation of the probe dynamics, including the state-dependent dipole force, radiation pressure, and photon-recoil diffusion. The simulated loss spectra
were fitted exactly as in the experiment, confirming this expectation. In contrast,
at the magic wavelength the resonance condition is position independent, so the
spectroscopy reflects the trap-center shift without any intensity-sampling
correction, and the power-dependent broadening from intensity sampling present at $1064$~nm is absent. The benchmark thus illustrates the ``mechanically clean'' advantage of the magic condition.

\section{Conclusions}

In summary, we have experimentally determined the $D1$ magic wavelength of $^{40}$K to be $\lambda_m = 1227.54(4)$~nm. This result was obtained by measuring the differential AC Stark shift in a wavelength-tunable optical tweezer, using loss spectroscopy on small atomic ensembles. The measured value is in excellent agreement with relativistic all-order calculations \cite{2025Portal}, and our total uncertainty of $0.04$~nm with comparable contributions from the wavelength calibration, its stability, and the zero-crossing statistics.

The significance of this result is better appreciated by considering our benchmark measurements at 1064.49~nm. While spectroscopy at standard trapping wavelengths must account for the thermal sampling of the inhomogeneous intensity distribution and for state-dependent forces during the probe pulse, the 1227~nm magic wavelength provides a
``mechanically clean'' environment. With balanced polarizabilities the resonance
is uniform across the trap, the line is narrow, and the spectroscopic signal
reflects the peak light shift without correction. This suppression of intensity-sampling systematics is essential for high-precision spectroscopy.

Beyond precision spectroscopy, operating at the magic wavelength provides immediate practical advantages for the control of $^{40}$K in optical tweezer arrays. It enables trap-independent fluorescence detection and facilitates the implementation of sub-Doppler cooling schemes, such as $D1$ gray molasses, directly inside the tweezers without the need for complex timing sequences or trap-off intervals. Furthermore, the elimination of differential shifts during light-assisted collisions will enhance the fidelity of deterministic single-atom loading. These capabilities provide a robust foundation for scalable quantum simulation and computation with neutral fermions, where high-fidelity state preparation and readout are paramount.

\begin{acknowledgments}
We thank Pavel Sidorenko and Yuval Shagam for their assistance with the laser frequency calibration. This research was supported by the Israel Science Foundation (ISF) under Grant No. 3348/25, the Pazy Research Foundation, and partially by the Helen Diller Quantum Center at the Technion.
\end{acknowledgments}

\section*{Data availability}
The data that support the findings of this article are available from the
authors upon reasonable request.

\end{document}